\documentclass[fleqn,twoside]{article}
\usepackage{espcrc2,epsf}

\def\be         {\begin{equation} }
\def\ee         {\end{equation} }
\def\bea                {\begin{eqnarray} }
\def\eea                {\end{eqnarray} }

\def\lsi{\raise0.3ex\hbox{$<$\kern-0.75em\raise-1.1ex\hbox{$\sim$}}}
\def\gsi{\raise0.3ex\hbox{$>$\kern-0.75em\raise-1.1ex\hbox{$\sim$}}}

\newcommand{\MeV}{\mathop{\rm MeV}}
\newcommand{\fm}{\mathop{\rm fm}}

\newlength{\figsize}
\newlength{\figoffset}
\newlength{\figbackup}
\newlength{\figendsp}

\def\preprints{
\vspace{-10ex}
{\small
\begin{tabbing}
\` Cambridge DAMTP-2001-92 \\
\` October 2001 \\
\end{tabbing} 
}
\vspace*{0.1in}
}

\title{
\preprints
Glueballs and topology with ${\cal O}(a)$--improved lattice QCD.}

\author{UKQCD and QCDSF Collaborations: A. Hart%
        \address{{\small DAMTP, Cambridge University, 
          Wilberforce Road, Cambridge CB3 0WA, UK.}}
        }

\begin{document}

\begin{abstract}
\noindent
We present evidence for unquenching effects in $N_f=2$, $16^3 32$
ensembles by comparing with `equivalent' quenched data at $\hat{r}_0
\simeq 5.0$. A (small) VEV for torelons signals (weak) string
breaking. A $15 - 20\%$ reduction in the scalar glueball mass relative to
quenched is argued to be (in part at least) a discretisation
effect. We find a chiral suppression of the topological susceptibility
consistent with expectations, and agreement between fermionic and
gluonic methods for measuring the topological charge.
\end{abstract}

\maketitle

To study the effects of vacuum polarisation in QCD, we should vary the
sea quark mass, $m_q$, having first removed discretisation
effects. With confident continuum extrapolations of many lattice
quantities unavailable, however, we must instead try to `freeze'
lattice artefacts. UKQCD does this by attempting to fix the lattice
Sommer scale, $\hat{r}_0 \simeq 5$, as $m_q$ is changed. Variations
due to residual changes in $\hat{r}_0$ are further reduced by the the
leading order discretisation errors for the ${\cal O}(a)$--improved
action being quadratic, rather than linear, in the lattice spacing,
$a$. An `equivalent' value of $\hat{r}_0$ is found in the quenched,
`gluodynamical' theory at coupling $\beta_{\rm Wil}=5.93$.  Details of
scales, parameters and matching can be found in
\cite{ukqcd01}.
We use as a measure of $m_q$ the square of the pseudoscalar `pion'
mass. Circumflexes denote dimensionless lattice quantities.

%
%
%
%
%

The failure of experiments to detect unambiguous glueball states is
believed to be a consequence of mixing between the light glueballs and
$q\bar{q}$ states. A major goal of lattice QCD is to predict the
mixing matrix elements, but the uncertainties in attempts
so far
\cite{toussaint00,michael99}
mean, however, that phenomenological attempts to describe the content
(gluonic or $q\bar{q}$) of the scalar sector glueball candidates lead
to widely differing results
\cite{close01}.

As a preliminary to a full mixing analysis
\cite{michael_prog},
we study the light glueballs and winding flux tubes on UKQCD dynamical
configurations and find, in common with
\cite{bali00},
that accurate mass estimates are possible on moderately sized
ensembles ($4000-8000$ HMC trajectories). Full results can be found
in
\cite{hart01},
along with a discussion of the applicability of the term `glueball'
when mixing is induced by vacuum polarisation.

On a spatial torus, Polyakov loops (PL) couple to colour flux tubes
that close upon themselves through the periodic boundary with unit
winding number. In the confining phase of gluodynamics such `torelons'
are stable, with the PL vacuum expectation value (VEV) zero even
after operator improvement. To a first approximation the flux tube
mass is the product of the string tension, $\hat{K}$, and
the spatial extent of the lattice, $L$.

\begin{figure}[t]

\leavevmode
\begin{center}

\hbox{%
\hspace{\figoffset}
\epsfxsize = \figsize
\epsffile{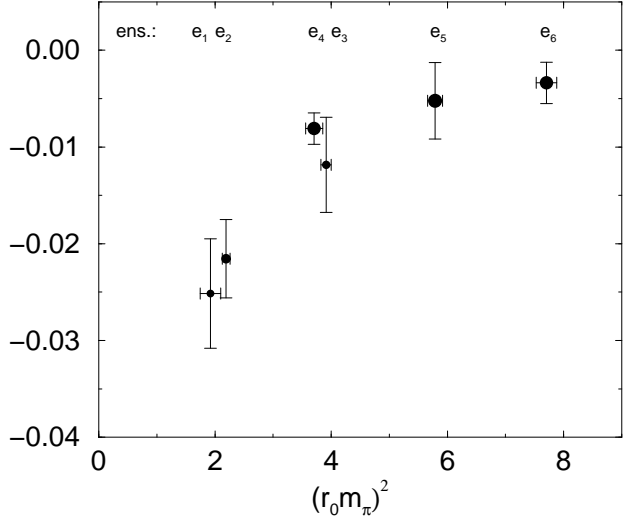}
}

\vspace{\figbackup}
\end{center}

\caption{The VEV of the spatial torelon.}
\label{fig_poly}

\vspace{\figendsp}
\end{figure}

Dynamical quarks break the centre symmetry that ensures the
orthogonality of contractible and non-contractible operators. The PL
can now have a non--zero VEV. This VEV should increase as $m_q
\rightarrow 0$ (and centre symmetry breaking gets worse) and, due to
the spatially periodic fermionic boundary conditions, be negative. In
a simple model it goes as $-(m_q)^{-L}$ for (moderately) heavy sea
quarks
\cite{hart01}.
This is `string breaking' of the confining flux tube, and
Fig.~\ref{fig_poly} confirms this picture for improved PL VEVs. With
the breaking of centre symmetry, mixings between torelons winding in
different directions becomes possible. Also allowed is mixing between
single (non-contractible) PL and (contractible) glueball
operators. This should be the leading source of (mass reducing) finite
volume effects in the scalar glueball: the double torelon that plays
this r\^{o}le in the quenched theory is too heavy to figure here. Both
mixings are very weak, however, as expected for a numerically small
VEV, and we can be confident that finite volume effects are small for
the scalar glueball. We can also thus use the torelon mass to obtain
$\hat{K}$, which agrees well with that from static interquark
potential fits
\cite{ukqcd01}. 
\begin{figure}[t]

\leavevmode
\begin{center}

\hbox{%
\hspace{\figoffset}
\epsfxsize = \figsize
\epsffile{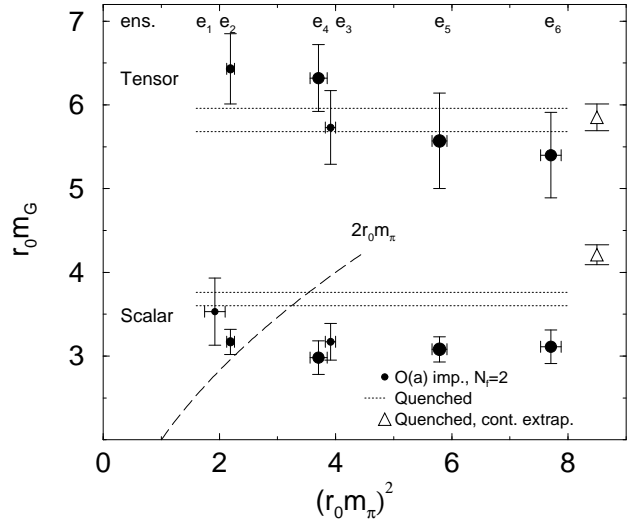}
}

\vspace{\figbackup}
\end{center}

\caption{The lightest glueball masses.}
\label{fig_gball}

\vspace{\figendsp}
\end{figure}

The masses of the lightest scalar and tensor states are shown in
Fig.~\ref{fig_gball}, with the `equivalent' quenched results at
$\beta_{\rm Wil}=5.93$, and the threshold for the scalar to decay to
pseudoscalars. Whilst the (admittedly noisy) tensor mass shows no
significant quenching effect, vacuum polarisation reduces the scalar
mass by $15 - 20\%$. The at most weak dependence of this figure on
$m_q$ and the $\pi \pi$ threshold suggest it is an increased
discretisation effect relative to the quenched theory, and that it may
not persist in the continuum limit. In gluodynamics, this `scalar dip'
is (probably) due to the scaling trajectory, $\beta_{\rm Wil} \to
\infty$, passing close to a critical point in the extended
fundamental--adjoint gauge coupling plane. The further reduction in
the presence of sea quarks for the ${\cal O}(a)$--improved action is
then not unexpected: hopping parameter expansions suggest the scaling
trajectory now passes even closer to this critical point. Measurements
of the couplings in such a mixed effective action appear to support
this. The lessened scalar mass at fixed, finite $a$ is thus not
necessarily indicative of a continuum effect
\cite{hart01a}.
\begin{figure}[t]

\leavevmode
\begin{center}

\hbox{%
\hspace{\figoffset}
\epsfxsize = \figsize
\epsffile{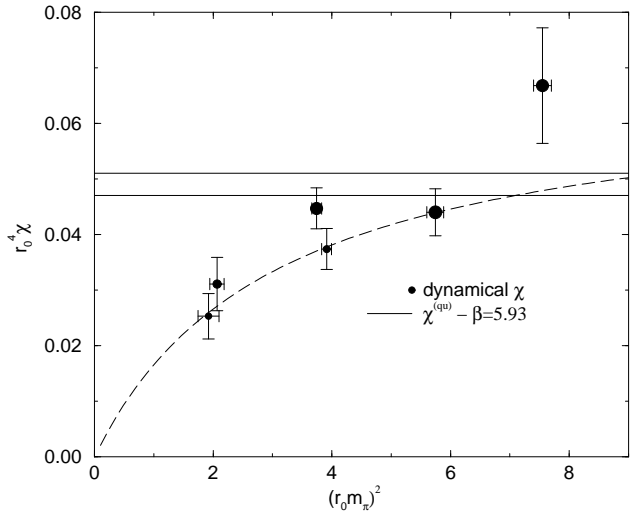}
}

\vspace{\figbackup}
\end{center}

\caption{The topological susceptibility.}
\label{fig_khi}

\vspace{\figendsp}
\end{figure}
%
%
%
%
%

%
%
%
%
%

The ability to vary
parameters fixed in Nature has made lattice Monte Carlo simulation a
valuable tool for investigating the r\^{o}le of topological
excitations in QCD and related theories. 

The topological susceptibility,
$
\chi = \langle Q^2 \rangle / V,
$
is the squared expectation value of the
topological charge,
\be
Q = \frac{1}{32\pi^2} \int d^4x 
\frac{1}{2} \varepsilon_{\mu \nu \sigma \tau}
F^a_{\mu \nu}(x) F^a_{\sigma \tau}(x)
\ee
normalised by the volume. Sea quarks induce an
instanton--anti-instanton attraction which becomes stronger in the
continuum limit, suppressing $\chi$
\cite{vecchia80}.
Given various assumptions
\cite{leutwyler92},
in the chiral limit of $N_f$ flavours we have
$
\chi = f_\pi^2 m_\pi^2/(2 N_f) + {\cal O}(m_\pi^4).
$
This is in a convention where the experimental value of the pion decay
constant $f_\pi \simeq 93 \ \MeV$, and requires $f_\pi^2 m_\pi^2 V \gg
1$, which holds for the lattices used here.
\begin{figure}[t]

\leavevmode
\begin{center}

\vspace*{1ex}
\hbox{%
\hspace{\figoffset}
\epsfxsize = \figsize
\epsffile{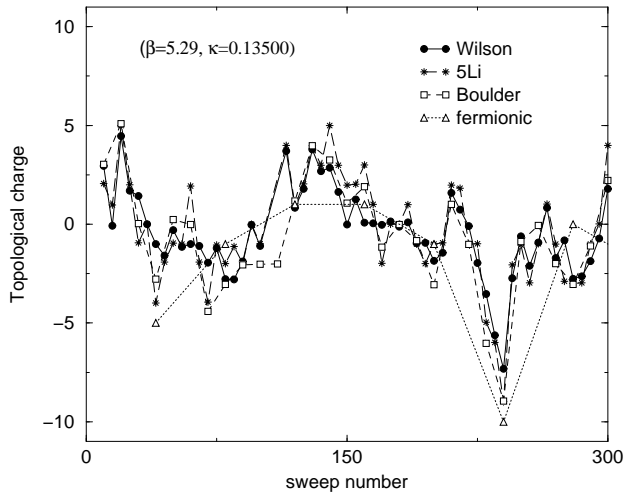}
}

\vspace{\figbackup}
\end{center}

\caption{Comparing topological charges.}
\label{fig_compare}

\vspace{\figendsp}
\end{figure}
The topological susceptibility is measured on UKQCD gauge fields using
a symmetrised `Wilsonian' operator, $\hat{Q}(n)$, having first
performed 10 Wilson action cools to regulate the ultraviolet
dislocations. Fig.~\ref{fig_khi} shows the chiral suppression in
$\hat{\chi}$ relative to the quenched limit with an interpolating fit
\cite{hart01b}. 
Including the $m_\pi^4$ term, we can fit the four most chiral data,
obtaining $f_\pi = 105 \ \pm 6 \ ^{+18}_{-10} \ \MeV$ at $a \simeq 0.1
\fm$. A fuller discussion of these results is available
in
\cite{hart01b,hart_other}.

%
%
%
%
%

The topology of the vacuum is clearly influenced by the sea quarks. At
finite $a$, however, $\hat{Q}$ is method dependent. To understand the
systematic biases this introduces, we compare several contrasting
methods with differing systematic errors for measuring $\hat{Q}$ on a
QCDSF ensemble for $a \simeq 0.1 \fm$. The first is as described
above. We also use the `5Li' method
\cite{forcrand97}, 
which is exact up to ${\cal O}(a^6)$ and approximately restores scale
invariance for large lattice instantons. We use the Boulder smearing
technique, which mimics a renormalisation group inspired smoothing of
the gauge fields
\cite{hasenfratz98}. 
Finally, we employ a fermionic charge operator on the `hot' gauge
fields
\cite{pleiter01}.

Fig.~\ref{fig_compare} shows Monte Carlo time series of the various
$\hat{Q}$, which have not been rounded to the nearest integer. We can
quantify the visually striking agreement through normalised
correlation functions using pairs of measurements. Preliminary results
give $85 - 95\%$ agreement between any pair of charge definitions;
subsequent work will compare systematic effects and the statistical
accuracy of topological susceptibility measurements. Other work in
this area can be found in
\cite{alles98}.

We thank Ph. de Forcrand for access to his code, and T. Kovacs and
D. Pleiter for results used in Fig.~\ref{fig_compare}.

{\small
%
%
%
%
%

}

\newpage
\vspace{5ex}
\noindent
{\bf Acknowledgments.}

\vspace{2ex}
\noindent
We thank the United Kingdom PPARC for support via the UKQCD research
and travel grants, and for postdoctoral funding.


\begin{thebibliography}{99}


\bibitem{ukqcd01}
C. Allton et al.,
hep-lat/0107021.

\bibitem{toussaint00}
D. Toussaint,
Nucl. Phys. (PS) 83-84 (2000) 151
[hep-lat/9909088].

\bibitem{michael99}
C. Michael et al.,
Nucl. Phys. (PS) 83-84 (2000) 185
[hep-lat/9909036];
C. McNeile, C. Michael,
Phys. Rev. D 63 (2001) 114503
[hep-lat/0010019].

\bibitem{close01}
F. Close, A. Kirk,
hep-ph/0103173.

\bibitem{michael_prog}
C. Michael, C. McNeile, A. Hart, M. Teper,
in progress.

\bibitem{bali00}
G. Bali et al.,
Phys. Rev. D 62 (2000) 054503
[hep-lat/0003012].

\bibitem{hart01}
A. Hart, M. Teper,
hep-lat/0108022.

\bibitem{hart01a}
A. Hart,
preprint DAMTP-2001-75.

\bibitem{hart01b} 
A. Hart, M. Teper, 
hep-lat/0108006.

\bibitem{hart_other} 
A. Hart, M. Teper, 
hep-ph/0004180;
hep-lat/0009008; 
Nucl. Phys. B (PS) 83-84 (2000) 476 [hep-lat/9909072].

\bibitem{vecchia80}
P. Di Vecchia, G. Veneziano, 
Nucl. Phys. B 171 (1980) 253.

\bibitem{leutwyler92}
H. Leutwyler, A. Smilga,
Phys. Rev. D 46 (1992) 5607.

\bibitem{forcrand97}
Ph. de Forcrand et al.,
Nucl. Phys. B 499 (1997) 409
[hep-lat/9701012].

\bibitem{hasenfratz98}
T. DeGrand, A. Hasenfratz, T. Kovacs, 
Nucl. Phys. B 520 (1998) 301
[hep-lat/9711032].

\bibitem{pleiter01}
D. Pleiter,
Proceedings of Lattice01.

\bibitem{alles98}
B. Alles et al.,
Phys. Rev. D 58 (1998) 071503
[hep-lat/9803008].

\end{thebibliography}
\end{document}